\documentstyle [12pt,epsf] {article} 
 
\catcode`@=12 
\begin{document} 
\vspace{0.5in} 
\oddsidemargin -.375in 
\newcount\sectionnumber 
\sectionnumber=0 
\def\be{\begin{equation}} 
\def\ee{\end{equation}} 
\thispagestyle{empty} 
\begin{flushright} 
UTPT-00-01 \\ 
January 2000\\ 
\end{flushright} 
\vspace {.5in} 
\begin{center} 
{\Large \bf{Effects Of Kaluza-Klein Excited $W$ 
\\ }} 
{\Large \bf{On Single Top Quark Production At Tevatron\\}} 

\vspace{.5in} 
{ \rm A. Datta{\footnote{email: datta@medb.physics.utoronto.ca}} 
and 
P.J. O'Donnell{\footnote{email: pat@medb.physics.utoronto.ca}} \\} 

{\it Department of Physics, 
University of Toronto\\ 
60 St George Street 
Toronto M5S 1A7, Canada\\} 

\vskip .5in 
{\rm Z.-H. Lin{\footnote{email: linzh@hptc5.ihep.ac.cn}} , 
X. Zhang{\footnote{email: xmzhang@hptc5.ihep.ac.cn}} 
and 
T. Huang{\footnote{email: 
huangt@hptc5.ihep.ac.cn}} \\} 

{\it CCAST (World Laboratory), P.O. Box 8730, Beijing 100080, P.R. China \\ 
\it and Inst. of High Energy Phys., Academia Sinica, P.O. Box 918(4), 
Beijing 100039, P.R. China 
\\} 
\vskip .5in 
\end{center} 
\begin{abstract} 
In extra dimension theories if 
the gauge bosons of the standard model propagate in the bulk of 
the extra dimensions 
then they will have Kaluza-Klein excitations that can 
couple to the standard model fermions. In this paper we study the effects 
of the first excited Kaluza-Klein mode of the $W$ on single top production 
at the Tevatron. We find that the cross section 
for the single top production can be significantly 
reduced if the mass of the first Kaluza-Klein excited 
$W \sim 1$ TeV. Hence, 
a measurement of the 
single top production cross section 
smaller than the standard model prediction 
would not necessarily imply 
$V_{tb} <1$ or evidence of extra generation(s) of fermions mixed with 
the third generation. 
\end {abstract} 
\newpage 
\baselineskip 24pt 

\section{\bf Introduction} 
An important issue in high energy physics is to understand the mechanism of 
mass generation. In the standard model, 
a fundamental complex Higgs scalar is introduced to break the electroweak 
symmetry and generate masses. However, 
arguments of triviality and naturalness 
suggest that the symmetry breaking sector of the 
standard model is just an effective theory. The top quark, 
with a mass of the order of the weak scale, is singled out to play a key role 
in probing 
the new physics beyond the 
standard model (SM) \cite{1}. 
The electroweak interactions of the top quark are particularly interesting and can 
be probed in the single top production and top decays. 
In this work we focus on the single top production at the Tevatron. 

Single top production at the Tevatron 
occurs within the SM in three 
different channels, the $s$-channel $W^*$ production, 
$q \bar q' \to W^* \to t \bar{b}$ 
\cite{schan, schan1, schan2, schanresum, boos,st},  
the 
$t$-channel $W$-exchange mode, $b q \to t q'$ 
\cite{boos,st, tchan, tchan1, tchan2, cpy, dougthesis, bordes, tchan3} 
(sometimes referred to as $W$-gluon 
fusion), and through $t W^{-}$ production \cite{tw}. 

The process $q{\overline q}\rightarrow t {\overline b}$, 
compared to the single top production via 
W-gluon fusion 
has the advantage that the cross section can be calculated reliably 
because the quark and 
antiquark structure functions at the relevant values of $x$ are 
better known than the gluon structure functions that enter in the 
calculation for the W-gluon cross section. Measurement of single top 
production cross section has been discussed in detail in 
Ref\cite{boos,st,Heinson}. In these references it is estimated that 
 single top production 
can be measured with an experimental error, at the one sigma level, of 
$\pm$ 19 \% 
at Run 2 (now called Run 2a) with an integrated luminosity of $2fb^{-1}$. 
The measured cross section can then be used 
to extract the CKM element $V_{tb}$ with a combined theoretical and experimental error of $\pm$ 
12-19\% in Run 2a depending on how one estimates the theoretical error. 
In Ref\cite{boos,st,Heinson} it was mentioned that there may be a 
Run 3 producing 
$30fb^{-1}$ of data and if only the 
$s$-channel $W^*$ production, 
$q \bar q' \to W^* \to t \bar{b}$ is used, then $V_{tb}$ could be 
extracted at Run 3 
with an error (including theoretical error) of about $\pm$ 5\%. At present 
Run 2a is expected to start next year and achieve ultimately 
an integrated luminosity of $2fb^{-1}$. The run beyond 
an integrated luminosity of $2fb^{-1}$ is no longer called Run 3 but is a 
continuation of Run 2 (Run 2b) and may achieve 
an integrated luminosity of $15fb^{-1}$ or higher. Update of the 
estimate on the precision in single top measurement at Run 2 since 
Ref\cite{boos,st,Heinson} is not yet available \cite{Heinson2}. As a rough estimate
for the errors in measuring $V_{tb}$ in Run 2b, operating at 
an integrated luminosity of $15fb^{-1}$, one can multiply the estimate
in ``Run 3'' presented in Ref\cite{boos,st,Heinson} by a factor
of $\sqrt{2}$.

The unitarity of the CKM matrix 
leads to a value of $V_{tb} \sim 1$. Hence a measurement of $V_{tb}$ 
which differs from 
unity would indicate presence of new physics. For instance a measurement of 
$V_{tb} <1$ is commonly taken to indicate
 the existence of new generation of fermions 
mixed with the third generation. 

Thus, it is possible that the effects of new physics will be revealed in single top production 
\cite{newphysics}. 
In this paper we consider 
effects that extra dimension theories can produce in single top production 
at the Tevatron. 
If in such theories, the gauge fields of the 
Standard Model(SM) live in the bulk of the extra dimensions 
then they will have Kaluza-Klein(KK) excitations. The possibility that 
the masses of the lowest lying of these states 
could be as low as 
$\sim$ a few TeV or less (of the order of the inverse size of 
the compactification radius ) leads 
to a very rich and exciting phenomenology at future and, possibly, existing 
colliders{\cite {old}}. 
Limits on the masses of the lowest lying excitations obtained from 
direct $Z'/W'$ and dijet bump searches at the Tevatron from Run 1 
indicate that 
they must lie above $\simeq 0.85$ TeV{\cite {tev}}. A null 
result for a search made with 
data from Run 2 will push this limit to $\simeq 1.1$ TeV . 
Model dependent limits can also be placed on the masses of the excitation 
from low energy observables and precision electroweak measurements 
\cite{edim,rw,lowenergy}. For instance in Ref\cite{edim} global fits to the 
electroweak observables, with certain assumptions, were found to provide 
lower bounds on the compactification scale, 
$M_c$, (which is equal to the mass of the first excited KK gauge boson) 
which were generically in the 2-5 TeV range depending on which standard model 
fermions, as well as the higgs boson, live in the bulk of the extra dimensions 
or are localized at different points of it. In fact Ref\cite{edim} found 
scenarios where global electroweak fits give a 95\% C.L upper and lower 
bounds on $M_c$ in the range 
$0.95 $ TeV $ \le M_c \le 3.44$ TeV. Note the analysis of 
Ref\cite{rw} assumed the standard model fermions to be stuck at the boundary of the extra dimension. 

In addition to the various assumptions, mentioned above, that are involved 
in putting bounds on $M_c$ from global electroweak fits there is another 
very important assumption made in all these analyses. In all these 
analyses it is 
assumed that the {\it only} new physics beyond the standard model arise 
from the the physics of the KK excitations of the standard model fields. For instance,
as mentioned in Ref\cite{rw}, in all these analyses the  gravity induced 
processes are assumed not to significantly
affect the electroweak observables. Note that, in general,
 the gravity induced processes
will affect electroweak observables, changing the bound on 
$M_c$ from electroweak data, but will not affect single
top production at tree level. In fact it is 
quite likely that there are additional new physics effects which may easily 
change the bounds on $M_c$ obtained from global fits to electroweak data. One 
 can represent the effects of this additional new physics in 
terms of higher dimensional operators in the effective Lagrangian framework. 
Recent studies have clearly demonstrated that the presence of higher 
dimensional operators can significantly effect global fits to the 
electroweak observables 
\cite{Hall}. In light of the above discussion we do not 
strictly enforce the bounds on $M_c$ from global electroweak 
fits in a specific model 
but rather assume that $M_c$ is in the same ballpark as obtained from global 
electroweak fits. In other words we assume that $M_c \sim $ TeV.

In this work we consider the contribution of the first excited KK mode 
of the $W$, denoted by $W_{KK}$, on the s-channel mode for the single top production 
at the Tevatron. This 
channel is more sensitive to the 
presence of a new charged resonance than the t-channel $W$-gluon fusion 
mechanism as was discussed in 
Ref\cite{topflavor}. This is 
because the momentum of the the s-channel resonance is time-like 
which leads to larger interference with the standard model amplitude than 
the t-channel process where the momentum of the $W_{KK}$ is space-like. For 
the 
s-channel process there can be a resonant 
enhancement of the amplitude which does not occur in the t-channel process. 
Note that the additional new physics effects, discussed above, which are 
not of gravitational origin may also affect 
single top production and has been extensively studied in Ref\cite{newphysics} 
and we do not consider these effects in this work. However, if the additional new physics is 
from gravity induced processes then there is no effect, at the tree level,
 on the s-channel mode for
 single top production which is mediated by the exchange of a charged boson. 

The paper is organized as follows. In section II, we calculate 
the effects of the 
excited Kaluza-Klein $W$ state on the single top production. 
In section III, we present our results and conclusions. 

\section{\bf Effect of KK excited $W$ in the single top production 
rate at Tevatron } 

To study the physics of the KK excited $W$ we use a model  
which is based on a simple extension of the 
SM to 5 dimensions (5D) \cite{edim,alex}. However, as discussed above, we
 do not assume that this model represents all the physics 
beyond the standard model. The 5D SM is probably a part of a more
fundamental underlying theory.

In the 5D SM  model the fifth dimension $x_5$ is compactified on the 
orbifold $S^1/{Z}_2$, a circle of radius $R$ 
with the identification $x_5\rightarrow -x_5$. 
This is a segment of length $\pi R$ with two 4D boundaries, one at 
$x_5=0$ and another at $x_5=\pi R$ 
(the two fixed points of the orbifold). 
The SM gauge fields live in the 5D bulk, 
while the SM fermions, $\psi$, and the Higgs doublets, 
can either live in the bulk 
or be localized on the 4D boundaries. We do not consider gravity in our 
analysis. It is possible that gravity might propagate in more 
extra dimensions than the SM fields. We do not expect gravity to 
affect single top production at the Tevatron. 

If the standard model fields live in the bulk then they 
will have KK excitations. 
The fields living in the bulk 
can be Fourier-expanded as 
\begin{eqnarray} 
\Phi_+(x_\mu,x_5)&=&\sum^{\infty}_{n=0} 
\cos\frac{nx_5}{R}\Phi^{(n)}_+(x_\mu)\, ,\nonumber\\ 
\Phi_-(x_\mu,x_5)&=&\sum^{\infty}_{n=1} 
\sin\frac{nx_5}{R}\Phi^{(n)}_-(x_\mu)\, , 
\label{fourier} 
\end{eqnarray} 
where $\Phi^{(n)}_{\pm}$ are the KK excitations of the 5D fields and the 
fields have been 
defined to be even or odd under the 
${Z}_2$-parity, i.e. $\Phi_{\pm}(x_5)=\pm\Phi_{\pm}(-x_5)$. 

As mentioned above the gauge fields live in the bulk. They are 
assumed to be even under the ${Z}_2$ parity. 
Their (massless) zero modes correspond to the 
standard model gauge fields. If the Higgs boson lives in the bulk then 
it is assumed to be even under the ${Z}_2$ parity also. 
Fermions in 5D have two chiralities, $\psi_L$ and $\psi_R$, that can 
transform as even or odd under the ${Z}_2$. The precise assignment 
is a matter of definition. It is assumed that 
$\psi_L$ ($\psi_R$) components of 
fermions $\psi$, which are doublets (singlets) 
under SU(2)$_L$ have even ${Z}_2$ parity and consequently only the $\psi_L$ 
of SU(2)$_L$ doublets and $\psi_R$ of SU(2)$_L$ singlets have zero modes. 
The fermions in this model couple to the KK excited gauge bosons only if they are localized 
on the 4D boundaries. The Lagrangian 
along with additional 
details and low energy phenomenology of this model can be found in \cite{edim,rw,alex} 
and will not be presented 
in this paper. 
For simplicity we will assume that there is one higgs doublet which along with the fermions 
that participate in the s-channel process for the single top production are localized on the 
4D boundary at 
$x_5=0$.

\begin{figure}[htb] 
\centerline{\epsfysize 6.2 truein \epsfbox{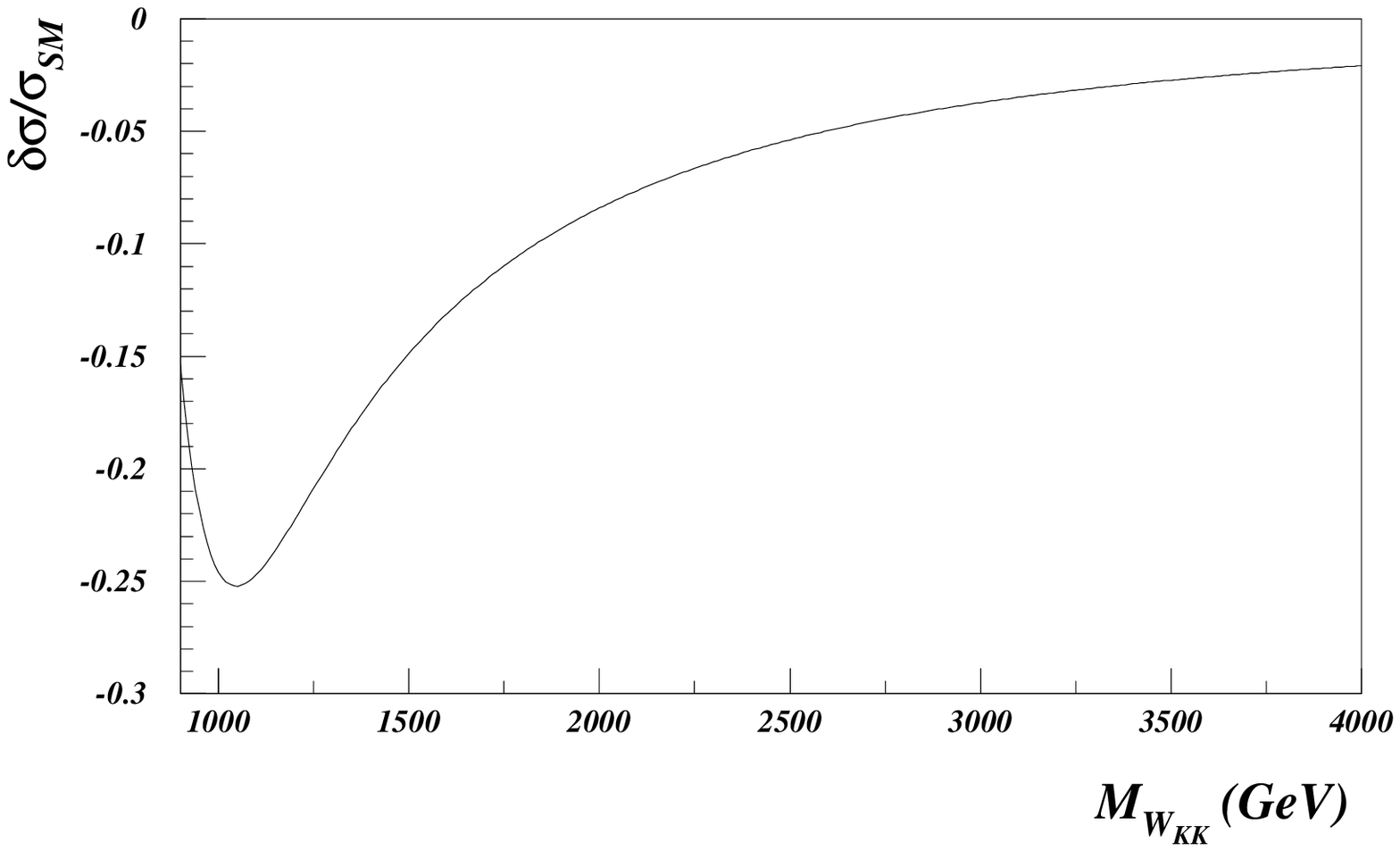}} 
\caption{$\Delta \sigma/\sigma_{SM}$ versus $M_{W_{KK}}$ the mass of the first KK excited 
$W$. } 
\end{figure} 

\begin{figure}[htb] 
\centerline{\epsfysize 6.2 truein \epsfbox{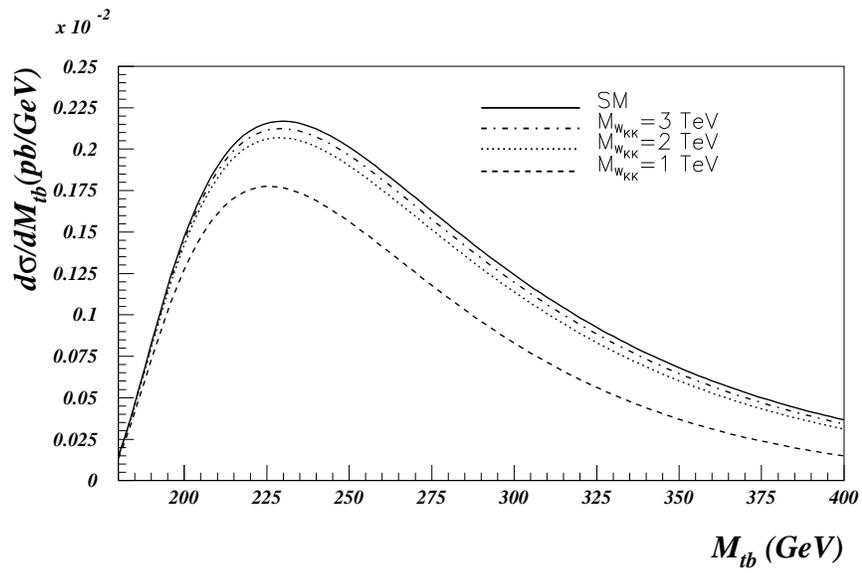}} 
\caption{The distribution $ \frac{d \sigma}{dM_{tb}}$ for various values of 
$M_{W_{KK}}$ for a limited range of $M_{tb}$. } 
\end{figure} 

\begin{figure}[htb] 
\centerline{\epsfysize 6.2 truein \epsfbox{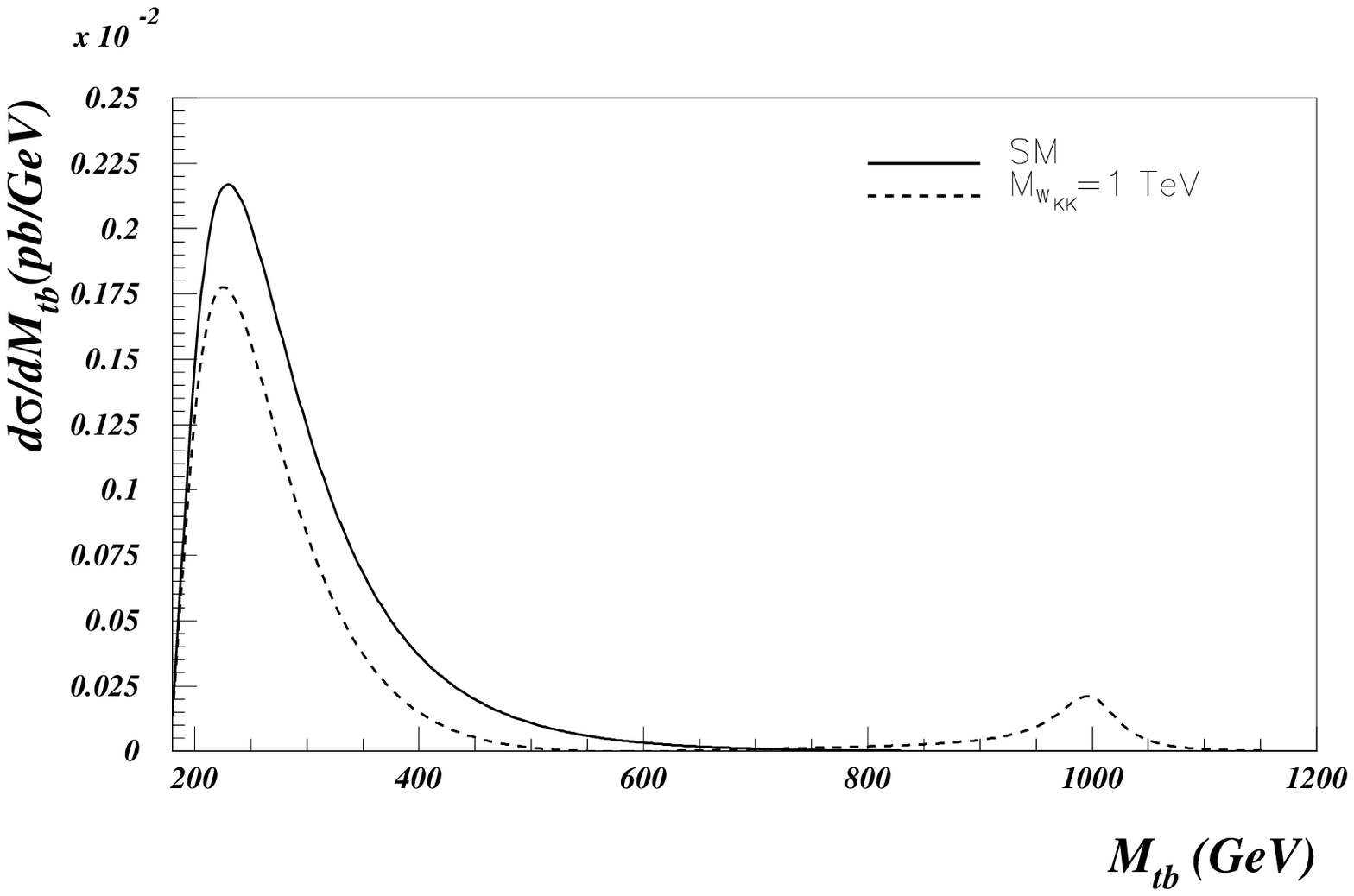}} 
\caption{The distribution $ \frac{d \sigma}{dM_{tb}}$ for 
$M_{W_{KK}} =1$ TeV for an extended range of $M_{tb}$. } 
\end{figure}

The effective four dimensional 
Lagrangian can be obtained after integrating over the fifth dimension . The 
piece of this Lagrangian relevant to our calculation 
is the charged electroweak sector and 
is given by

\begin{equation} 
{\cal{L}}^{ch}=\sum_{a=1}^2 {\cal{L}}_a^{ch} +{\cal{L}}_{new}
\end{equation} 
with 
\begin{eqnarray} 
{\cal{L}}_{a}^{ch}&=& 
\frac{1}{2}m^{2}_W 
W_a\cdot W_a 
+\frac{1}{2}M_c^2\sum_{n=1}^\infty \, n^2 \, W_a^{(n)}\cdot W_a^{(n)} 
\nonumber\\ 
&-&g\,W_a\cdot J_a-g\,\sqrt{2} J^{KK}_a\cdot \sum_{n=1}^\infty W_a^{(n)}\, , 
\end{eqnarray} 
where $m^{2}_W=g^2v^2/2$, the weak angle $\theta$ is defined by 
$e=g\, s_\theta=g'\, c_\theta$, while the currents are 
\begin{eqnarray} 
\label{currents} 
J_{a\mu}&=&\sum_\psi \bar{\psi}_L \gamma_\mu \frac{\sigma_a}{2}\psi_L\, , 
\nonumber\\ 
J_{a\mu}^{KK}&=& \sum_\psi \varepsilon^{\psi_L}\bar{\psi}_L \gamma_\mu 
\frac{\sigma_a}{2}\psi_L\, . 
\end{eqnarray} 
Here $\varepsilon^{\psi_L}$ takes the value 
1(0) for the $\psi_L$ living in 
the boundary(bulk). The mass of the $n^{th}$ excited KK state of the 
$W$ is given by $nM_c=n/R$ where R is the compactification radius. In this work we 
consider only the $n=1$ state. The term $ {\cal{L}}_{new}$ represents the additional new
physics beyond the 5 dimensional standard model the structure of which remains
unknown till the full underlying theory is understood. The coupling of
KK excited $W$ to the standard model is determined in terms of the
Fermi coupling, $G_F$, up to corrections of $O(m_Z^2/M_c^2)$ \cite{edim,rw} . For $M_c \sim $ TeV
the $O(m_Z^2/M_c^2)$ effects
 are small for single top production and therefore we do not include
these effects in our calculations. We have  ignored the mixing of the
$W$ with $W_{KK}$ which is also an $O(m_Z^2/M_c^2)$ effect. Thus, assuming
 the $W_{KK}$
decays only to standard model particles, the predicted effect of $W_{KK}$
on single top production depends, in addition to the SM parameters, only on 
the unknown
mass of the $W_{KK}$.

The cross section for $p {\overline p}\rightarrow t{\overline b} X$ 
is given by 
\begin{eqnarray} 
\sigma(p {\overline p}\rightarrow t{\overline b} X) & = & 
\int dx_1dx_2[u(x_1){\overline d}(x_2)+ u(x_2){\overline d}(x_1)] 
\sigma(u{\overline d}\rightarrow t{\overline b}) . \ 
\end{eqnarray} 
Here $u(x_i)$, ${\overline d}(x_i)$ are the $ u $ and the ${\overline 
d}$ structure functions, 
$x_1$ and $x_2$ are the parton momentum fractions and the indices 
$i=1$ and $i=2$ refer to the proton and the antiproton. 
The cross section for the process 
$$u(p_1) + {\overline d}(p_2) \rightarrow W^* \rightarrow {\overline 
b}(p_3) + t(p_4) , $$ 
is given by 
\begin{eqnarray} 
\sigma & = & \sigma_{SM}\left[1 + 4 \frac{A}{D} +4 \frac{C}{D} \right], \nonumber\\ 
A & =& (s-M_W^2)(s-M_{W_{KK}}^2) + M_WM_{W_{KK}}\Gamma_{W}\Gamma_{W_{KK}}, \nonumber\\ 
C & = & (s-M_W^2)^2 +(M_W\Gamma_{W})^2, \nonumber\\ 
D & = & (s-M_{W_{KK}}^2)^2 +(M_{W_{KK}}\Gamma_{W_{KK}})^2, \ 
\end{eqnarray} 
and 
\begin{eqnarray} 
\sigma_{SM} & = & \frac{g^4}{384 \pi} \frac{(2s+M_t^2)(s-M_t^2)^2}{s^2[ 
(s-M_W^2)^2 +(M_W\Gamma_{W})^2]}. \ 
\end{eqnarray} 
Here $s=x_1x_2S$ is the parton center of mass energy while 
$S$ is the $ p{\bar p}$ center of mass energy . To calculate the width 
of the $W_{KK}$ we will assume that it 
decays only to the standard model particles. The $W_{KK}$ will then have the same 
decays as the $W$ boson but in addition it can 
also decay to a top-bottom pair which is kinematically forbidden for the $W$ boson. 
The width of the $W_{KK}$, $\Gamma_{W_{KK}}$, is then given by 
\begin{eqnarray} 
\Gamma_{W_{KK}} & \approx & 
\frac{2M_{W_{KK}}}{M_{W}}\Gamma_W+ 
\frac{2M_{W_{KK}}}{3M_{W}}\Gamma_W\cdot X ,\nonumber\\ 
X & = & ( 1-\frac{M_t^2}{M_{W_{KK}}^2}) 
( 1-\frac{M_t^2}{2M_{W_{KK}}^2} - \frac{M_t^4}{M_{W_{KK}}^4}). \ 
\end{eqnarray} 
where $\Gamma_W$ is the width of the $W$ boson and we have neglected the 
mass of the $b$ quark along with the masses of the lighter quarks and the 
leptons. 
\section{Results} 

In Fig. 1, we plot $\Delta \sigma/\sigma$ versus $M_{W_{KK}}$ , the mass of 
the first excited KK $W$ state, where $\Delta \sigma$ 
is the change in the 
single top production cross section in the presence of $W_{KK}$ 
and $\sigma$ is the standard 
model cross section\footnote{ We have 
not included the QCD and Yukawa corrections to the single top quark 
production rate. They will enhance the total rate, but not change 
the 
percentage of the correction of new physics to the cross section. 
}. 
We have used the CTEQ \cite{CTEQ} structure 
functions for our calculations and obtain a standard model cross section 
of 0.30 pb for the process $p {\overline p}\rightarrow t{\overline b} X$ 
at $\sqrt{S}= 2$TeV. 
We observe from Fig. 1 that the presence of $W_{KK}$ can 
lower the cross section by as much as 25 \% for $M_{W_{KK}} \sim 1 $TeV. 
This has an important implication for the measurement of $V_{tb}$ using 
the s-channel mode at the Tevatron. It was pointed out in Ref\cite{topflavor}
that there could be models where
 the presence of an additional $W$( denoted as
$W'$) could lead to a measurement of the 
cross section for the s-channel $p {\overline p}\rightarrow t{\overline b} X$ 
smaller than the standard model prediction. A specific example of such a model with a $W'$  that causes a significant 
decrease of the single top cross section can be found in Ref\cite{ES}.  
This could, as pointed out in 
Ref \cite{topflavor},
lead one to conclude that
$V_{tb} <1$ which could then be wrongly interpreted as 
evidence for the existence of new generation(s) of fermions mixed with the 
third generation. Our work provides another specific example of such a model
and our results clearly demonstrates that
a measurement of the 
cross section for the s-channel $p {\overline p}\rightarrow t{\overline b} X$ 
smaller than the standard model prediction 
would not necessarily imply 
$V_{tb} <1$ or evidence of extra generation(s) of fermions mixed with 
the third generation. Note that, as mentioned above, the predicted 
effect of $W_{KK}$
on single top production depends, in addition to the SM parameters,
 only on  the unknown 
mass of the $W_{KK}$, while in 
most other $W'$ models the predictions for single top production depend, 
in addition to the SM parameters, 
on the unknown mass of the $W'$ as well as on 
 unknown mixing parameter(s). 

 Note that the $W_{KK}$ can also be searched 
at the Tevatron through its decay into a high energy lepton and a neutrino 
if it couples to the leptons. Searches for this resonance at the Tevatron 
allow discovery at 1.11 TeV and 1.34 TeV with 2 $fb^{-1}$ and 
20 $fb^{-1}$ for $\sqrt{S}=2$ TeV \cite{rw}. In this energy range, 
as shown in Fig. 1, there will be significant effects on the single top production rate. 
If the SM leptons are allowed to live in the bulk then they will not couple to 
$W_{KK}$ and so it is no longer possible 
to search for this resonance through its decay to leptons. In such a scenario, 
single top production could be a very effective 
probe of the $W_{KK}$ resonance. 

In Fig.2 we show the $t{\bar b}$ invariant mass distribution $\frac{d\sigma}{d M_{tb}}$, where 
$M_{tb}$ is the invariant mass of the $t{\bar b}$ pair for various values of $M_{W_{KK}}$. 
We see a significant 
decrease in the signal at $M_{W_{KK}} \sim 1$ TeV for lower values of $M_{tb}$. 
For lower $M_{tb}$, the interference term ($\frac{A}{D}$) in Eq. (6) has a stronger 
effect 
than the direct term ($\frac{C}{D}$) which leads to a reduction of the signal. 
In Fig. 3 we show 
the $t{\bar b}$ invariant mass distribution $\frac{d\sigma}{d M_{tb}}$ for 
an extended range of $M_{tb}$ for $M_{W_{KK}}=1$TeV. As we go 
to larger values of $M_{tb}$, close to the resonance region, the direct term in Eq. (6) 
becomes dominant. This leads to a bump 
in the resonance region. However, the signal is considerably reduced because of smaller 
parton distributions. 

In conclusion, we have studied the effects of a KK excited $W$ 
on the cross section of the single top production at the Tevatron. 
The model of $W_{KK}$ considered in this work
leads to a definite structure for its coupling to the standard model fields.
 Moreover,
the prediction for single top production,
 up to very small corrections, depend only on one additional unknown parameter, the mass of the $W_{KK}$. This is unlike the usual $W'$ models, which 
 require extending the 
standard model gauge group and the predictions for single top production 
 depend 
on the unknown mass of the $W'$ as well as on 
additional unknown mixing parameter(s). 
Our results show 
that the cross section for s channel single top quark production 
can be significantly 
reduced, by about 25\%, from the standard model value for $M_{W_{KK}} \sim 1$TeV. Therefore
the s channel single top production could be a very effective 
probe of the $W_{KK}$ resonance with a mass $\sim$ TeV at the Tevatron.

{\bf Acknowledgment:} 
This work was supported in part by Natural Sciences and Engineering Research Council 
of Canada(A. Datta and P.J. O'Donnell) 
and by Chinese National Science Foundation (T. Huang, X. Zhang and Z.-H. Lin). 
We thank A. P. Heinson, W.-B. Lin and S.-H. Zhu for discussions.

\end{document}